**To the problem of finding the transmission and reflection bands of a plane electromagnetic wave falling on an ideal structure**

**A.Zh. Khachatrian**

*State Engineering University of Armenia, 375049,Terian 109, Yerevan, Armenia*
akhachat@www.physdep.r.am

In the present paper we investigate the transmission and reflection band behavior for a plane electromagnetic wave falling obliquely on an ideal layered structure. The dependence of this behavior on the problem parameters and wave incident angle is considered. It is shown, that in general case the band width is a non-monotonous function of the problem parameters. A condition is found, which defines the possibility of the contact of the transmission bands. This condition has the same form for $s$ and $p$ waves. It is also shown that irrespective of the wave polarization, the transmission coefficient equals to the unit at the contact points.

PACS: 78.20. Bh; 78.20.B

Keywords: one-dimensional periodic structure, transmission band, $s$ and $p$ waves.

Recently the theory of quasi-particles to describe the materials having artificial translational symmetry has been energetically developed. This theory is very important for the general wave theory and also for many physical applications [1-5]. The important features of these structures due to their microscopic influence upon the quasi-particles (electrons, phonons, photons, etc.) are that the particles energy spectrum gets the band nature, and the energy itself becomes a function of the quasi-momentum. On the other hand, the periodic system properties serve as the basis to explain a great deal of physical phenomena. For instance, one could mention the results of Bloch on solution of the differential partial equations, which were taken as a basis for the electronic theory of solid crystal state by Brillouen.

At present it is possible to make multilayer and periodic structures, using the achievements of the modern technology mostly connected with new materials syntheses by methods of molecular beam epitaxy. The parameters of these structures vary in a rather wide range. Due to this, a tendency prevails in solid state electronics and optoelectronics, when superstructure are used as working elements when constructing such devices as filters, modulators, sensors, etc. Such working elements endow these devices essentially new characteristic, which are absent in the devices whit homogeneous working elements [6-14].

In the present paper we investigate the problem of transmission of a plane electromagnetic wave obliquely falling on an ideal periodic structure. In particular, we study the transmission and reflection bands behavior for a structure, which is a homogeneous layer periodically repeated in space. We try to find the bands dependencies on the structures parameters, such as the layer refractive index, the layer thickness, the distance between the layers, and the incident angle. In our opinion, such formulation of the problem is of both physical and practical interest. In particular, it allows observing dynamics of changes the system transmission when one of its parameter varies. And such formulation is also connected with the necessity of prediction of structures with predetermined spectral characteristics.

The problem considered below is well known, and its solution is found in many academic books (e.g. [15]). So below we limit our selves only to the problem stating and final results. The remaining part of the paper will contains the discussion of the results.

Let us consider a plane monochromatic electromagnetic wave obliquely falling on an ideal layered structure. The dielectric constant of the considered structure can be represented in the form:

$$\varepsilon(z) = \begin{cases} \varepsilon_0, & z < 0, \\ \cdots & \cdots \quad \cdots \quad \cdots \\ \varepsilon, & (n-1)a < z < (n-1)a + d, \\ \varepsilon_0, & (n-1)a + d < z < na, \\ \cdots & \cdots \quad \cdots \quad \cdots \\ \varepsilon_0, & z > L, \end{cases} \quad (1)$$

where $n = 1, 2, \cdots, N-1$ ( $N$ is the number of layers), $a$ is the structure period, $\varepsilon_0$ and $\varepsilon$ are the dielectric constants in the mentioned regions, $d$ is the layer thickness and $L = Na - b$. We take the magnetic permeability equal to the unit ($\mu = 1$) everywhere.

Let the unit basis vectors are $\vec{e}_1$, $\vec{e}_2$, $\vec{e}_3$. And let the wave vector of the filling wave is in the plane $(x, z)$. Then, taking the time factor as $\vec{E} \exp\{-i\omega t\}$, we write the spatial dependence of intensity $\vec{E}$ as:

$$\vec{E}(\vec{r}) = \begin{cases} \vec{E}_0 \exp\{i\vec{k}_0\vec{r}\} + \vec{E}_r \exp\{i\vec{k}_r \vec{r}\}, & z < 0, \\ \vec{E}_t \exp\{i\vec{k}_0\vec{r}\}\}, & x > L, \end{cases} \quad (2)$$

where $|\vec{k}_0| = |\vec{k}_r| = k_0 = \omega\sqrt{\varepsilon_0}/c$ and

$$\vec{k}_0 = k_0 \sin\alpha\, \vec{e}_1 + k_0 \cos\alpha\, \vec{e}_3, \quad \vec{k}_r = k_0 \sin\alpha\, \vec{e}_1 - k_0 \cos\alpha\, \vec{e}_3. \quad (3)$$

The transmission coefficients $|T_N^s|^2$, $|T_N^p|^2$ for $s$ and $p$ waves in the case of structure (1) can be written in the form [15] (see also [16]):

$$\left|T_N^{s,p}\right|^2 = \frac{\sin^2\beta^{s,p}/\sin^2 N\beta^{s,p}}{\left|r^{s,p}/t^{s,p}\right|^2 + \sin^2\beta^{s,p}/\sin^2 N\beta^{s,p}}, \quad (4)$$

where the index $s$ means polarization in the direction perpendicular to the plane $(x, z)$ and $p$ denotes the polarization in the plane $(x, z)$. In (4) the following designations are introduced: $r^{s,p}$

and $t^{s,p}$ are the reflection and transmission amplitudes of $s$ and $p$ waves respective to one layer only. The parameter $\beta^{s,p}$, which can be both real and imaginary, determines the nature of wave transmission through the periodic structure: $\cos\beta^{s,p} = \text{Re}(\exp\{-ik_0 a\cos\alpha\}/t^{s,p})$. When $\beta^{s,p}$ is real transmission takes place for an arbitrary $N$. When $\beta^{s,p}$ is imaginary, the increase of $N$ leads to the total wave reflection. These parameters $\beta^{s,p}$ are defined as follows:

$$\cos\beta^s = \cos\{k_{0z}b\}\cos\{k_z d\} - \frac{k_{0z}^2 + k_z^2}{2k_{0z}k_z}\sin\{k_{0z}b\}\sin\{k_z d\}, \qquad (5)$$

$$\cos\beta^p = \cos\{k_{0z}b\}\cos\{k_z d\} - \frac{(\varepsilon/\varepsilon_0)k_{0z}^2 + (\varepsilon_0/\varepsilon)k_z^2}{2k_{0z}k_z}\sin\{k_{0z}b\}\sin\{k_z d\}. \qquad (6)$$

In the expressions (5), (6) the following designations are introduced: $b = a - d$, $k_{0z} = k_0\cos\alpha$ and $k_z = \omega\cos\gamma\sqrt{\varepsilon}/c$. The angles $\alpha$ and $\gamma$ connected by refraction law $\sqrt{\varepsilon_0}\sin\alpha = \sqrt{\varepsilon}\sin\gamma$. Therefore, when $\alpha$ is larger than the critical angle of the total reflection ($\alpha' = \arcsin\sqrt{\varepsilon/\varepsilon_0}$), $k_z$ is to be replaced by $k_z = iq_z$ ($q_z$ - real). The latter means the expressions (5), (6) are to be analytically continued.

According to (4)-(6) the equations, defining the even and odd reflection band borders are as follows:

$$\cos\beta^{s,p} = 1 \text{ ànd } \cos\beta^{s,p} = -1. \qquad (7)$$

It can be shown, that both for $s$ and $p$ waves each of equations (7) gives two independent equations. According to (5), (7), the equations determining the reflection bands for $s$ wave can be written as:

$$tg(k_{0z}b/2) = -\frac{k_{0z}}{k_z}tg(k_z d/2), \; tg(k_{0z}b/2) = -\frac{k_z}{k_{0z}}tg(k_z d/2), \qquad (8)$$

$$tg(k_{0z}b/2) = \frac{k_{0z}}{k_z}ctg(k_z d/2), \; tg(k_{0z}b/2) = \frac{k_z}{k_{0z}}ctg(k_z d/2). \qquad (9)$$

For the reflection bands of $p$ waves, from equations (7) and (6) we have the following equations:

$$tg(k_{0z}b/2) = -\frac{\varepsilon k_{0z}}{\varepsilon_0 k_z}tg(k_z d/2), \; tg(k_{0z}b/2) = -\frac{\varepsilon_0 k_z}{\varepsilon k_{0z}}tg(k_z d/2), \qquad (10)$$

$$tg(k_{0z}b/2) = \frac{\varepsilon k_{0z}}{\varepsilon_0 k_z}ctg(k_z d/2), \; tg(k_{0z}b/2) = \frac{\varepsilon_0 k_z}{\varepsilon k_{0z}}ctg(k_z d/2). \qquad (11)$$

When the wave incidence angle exceeds the critical angle of total reflection, then equations(8), (9) take the form

$$tg(k_{0z}b/2) = -\frac{k_{0z}}{q_z}th(q_z d/2), \quad tg(k_{0z}b/2) = \frac{q_z}{k_{0z}}th(q_z d/2), \qquad (12)$$

$$tg(k_{0z}b/2) = -\frac{k_{0z}}{q_z}cth(q_z d/2), \quad tg(k_{0z}b/2) = \frac{q_z}{k_{0z}}cth(q_z d/2). \qquad (13)$$

Similarly, after $k_z = iq_z$ replacement, we have for equations (10), (11)

$$tg(k_{0z}b/2) = -\frac{\varepsilon k_{0z}}{\varepsilon_0 q_z}th(q_z d/2), \quad tg(k_{0z}b/2) = \frac{\varepsilon_0 q_z}{\varepsilon k_{0z}}th(q_z d/2), \qquad (14)$$

$$tg(k_{0z}b/2) = -\frac{\varepsilon k_{0z}}{\varepsilon_0 q_z}cth(q_z d/2), \quad tg(k_{0z}b/2) = \frac{\varepsilon_0 q_z}{\varepsilon_0 k_{0z}}cth(q_z d/2). \qquad (15)$$

It is interesting to mention, that for values of the incidence angle larger than the total reflection one, when the layer width $d$ increases ($q_z d \gg 1$) the equations that determine borders of odd and even bands of reflection, both for case of $s$ and $p$ waves, turn into each other. Indeed, in equations (12) - (15) if $q_z d \gg 1$, it is possible to put $th(q_z d/2) \approx cth(q_z d/2) \approx 1$. Therefore in this case four of equations (12), (13) and (14), (15) turn into the following couples of equations:

$$tg(k_{0z}b/2) = -\frac{k_{0z}}{q_z}, \quad tg(k_{0z}b/2) = \frac{q_z}{k_{0z}}, \qquad (16)$$

and

$$tg(k_{0z}b/2) = -\frac{\varepsilon k_{0z}}{\varepsilon_0 q_z}, \quad tg(k_{0z}b/2) = \frac{\varepsilon_0 q_z}{\varepsilon k_{0z}}. \qquad (17)$$

Equations (16), (17) define the spectrum of antisymmetric and symmetrics mode of $s$ and $p$ polarized waves, which propagate in a wave-guide regime inside a homogeneous layer with a dielectric constant $\varepsilon_0$ and layer thickness $b$.

The $\varepsilon_0$ layer is put into an $\varepsilon$ medium. This result is clear if one takes into account that the condition $|\cos\beta^{s,p}| \leq 1$ defines not only the wave transmission bands, but also gives the region of wave frequencies able to exist without a dimping inside the ideal structure infinite in both sides. In the total reflection regime the wave energy mainly is concentrated in optically more dense layers. When the width of the layers increases, the field energy, which is concentrated inside the optically less dense regions decreases. Due to this fact, the field inside the infinite structure represents itself an infinite number of separate waves, everyone of which propagate in wave-guide regime inside the optically more dense layers.

Below we bring the results of numerical calculations, which show band boarders dynamic dependence on the problem parameters . On the fig.1-3 the calculation was carried out, assuming that $\varepsilon_0$ medium is optically less dense than that of $\varepsilon$ ($\varepsilon_0 < \varepsilon$ and $\varepsilon_0 = 1$). Figure 1 shows the transmission and reflection bands locations, depending on the layer dielectric constant at normal incidence for eight different values of dimensionless parameter $b/d$. As it is seen from the figure, the increase of $\varepsilon$ shifts both transmission and reflection bands to the lower frequency region. From the given graphs it also follows that when the distance between the layers becomes larger compared to their thickness the band widths have more tendency of decreasing. For instance, if $\varepsilon = 10$ twelve allowed bands are located inside the considered frequency region ($0 < \omega d / c < 10$), at $b = 0.5d$; and at $b = 4d$ the number of the bands is twenty three in the mentioned region. It is to be mentioned that on all figures some band contact effects can be observed for certain values of $\varepsilon$, and depending on the band locations and value of $b/d$ for one forbidden band, the contacts of its borders can take place one or more times. On figure 2, the band locations are represented depending on the dimensionless parameter $d/b$ for eight different values of layer dielectric constant. As it is seen from the figures, the increase of $d/b$ (as in the case of increase of $\varepsilon$ (see figure 1)), leads to the shift of the transmission and reflection bands to the lower frequency region. It is also to be mention, that in all the considered cases of figure 2 the band contact effect is present (see below). In the figure 3 the transmission and reflection band locations are shown for $s$ and $p$ waves, depending on $d/b$ for four different values of the incidence angle $\alpha$, for $\varepsilon = 7$. It is to be mentioned that on figure 3 the band contact effects are observed in four cases only, in contrast with the cases considered on figure 2.

On figures 4 and 5 the optically more dense cases of $\varepsilon_0$ are brought ($\varepsilon < \varepsilon_0$ and $\varepsilon_0 = 6$). The locations of bands for $s$ and $p$ waves (see figures 4 and 5, respectively) are plotted, depending on the layer dielectric constant for two values of $b/d$, and at four different incidence angles $\alpha$. As it is seen from the graphs, the reflection bands decrease, when the dielectric constant, beginning at its certain value, increases, and the allowed lines, composing to the allowed bands, correspond to the wave-guide layers modes (see above). When $\varepsilon$ tends to $\varepsilon_0$ all the transmission bands merge to single band in all the considered cases. Indeed, when $\varepsilon = \varepsilon_0$ the layered system represents itself an infinite homogeneous medium, in which a wave of an arbitrary frequency and polarization can freely propagate.

Now we present some conclusions, which can be made discussing the material of figures 1-5. First of all, one can state that transmission band width dependence on the layer dielectric constant and its thickness, etc., is not a monotonous function. This non-monotony appears at those values of the problem parameters, at which the forbidden band width vanishes. In the case of an optically more dense layer ($\varepsilon > \varepsilon_0$) the increment $\varepsilon - \varepsilon_0$ leads to a shift of the transmission bands to the lower frequency region, and decreases the bands width, after all. . In the case of an optically less dense layer ($\varepsilon < \varepsilon_0$) the increment $\varepsilon_0 - \varepsilon$ shifts the transmission bands to the higher frequency region. When the incriminate $\varepsilon_0 - \varepsilon$ is the highest the upper transmission bands are changed into lines, which correspond to those wave frequencies, that propagate inside the optically more dense layers in wave-guide regime. It is to be noted that then the incident angle $\alpha$ is larger, than the less increment $\varepsilon_0 - \varepsilon$ is needed to convert the transmission lines into the bands.

Finally, let us consider the contact effect of transmission bands. It is present in practically in all figures 1-5, at many points. First of all, let us note that a contact means that the forbidden band width is zero, i.e. its boarders are contracted into a point. From the four equations brought above ((8), (9) for $s$ wave and (10), (11) for $p$ wave) every separate equation can define both the upper and lower band borders, depending on the problem parameters values. It is to be noted, that if an equation defines the lower (upper) forbidden band boarder on the left hand side of a contact point, then the same equation defines the upper (lower) boarder on the right hand side of the contact point.

The condition of a forbidden band width being zero means, that the equations, defining different band boarders, give the same solution for certain parameters of the problem. It follows from equations (8), (9) and (10), (11), that both for $s$ and $p$ waves the mentioned condition has the same form. This condition is represented by two equations:

$$k_{0z}b = n\pi \text{ and } k_z d = m\pi, \qquad (18)$$

where $n$, $m$ are integers. To require (18) means to require the well known Fabry-Perot interferometer condition in the both layers simultaneously.

In particular, it follows from (18), that the band contact is possible only if the problem parameters are connected by the following expression:

$$\frac{\cos\alpha}{\sqrt{\varepsilon/\varepsilon_0 - \sin^2\alpha}} = \frac{d}{b}\frac{n}{m}, \qquad (19)$$

From the first equation of (18) it follows, that the contact of bands takes place at frequencies:

$$\omega = \frac{\pi n c}{b\sqrt{\varepsilon_0}\cos\{\alpha\}}. \tag{20}$$

Note, that at the given parameters of periodic structure $\cos\alpha$ in (20) is to be determined from (19). It is to be noted, that if both $n$ and $m$ are even or add simultaneously, then equations (8) and (10) have the same solution. Otherwise, the equations (9) and (11) have the same solution.

To understand the physical consequences of allowed bands contact effect let us consider the transmission coefficient (4) behavior at the contact point. As it is well known, from the theory of wave propagation in periodic systems, the system length increase at band boarders leads to the total reflection. For the considered structure the transmission coefficient at the allowed band boarders has the form [15]:

$$\left|T_N^{s,p}\right|^2 = \frac{1}{1+N^2\left|r^{s,p}/t^{s,p}\right|^2}. \tag{21}$$

Expression (21) obviousely tends to zero when $N$ increases, if only at the allowed band boarder $r^{s,p} \neq 0$. If $r^{s,p} = 0$, the transmission coefficient is equal to the unit for an arbitrary $N$. Let us remember, the reflection amplitudes vanish for both $s$ and $p$ waves if $k_z d = m\pi$ ($r^{s,p}=0$, see for example [16]). On the other hand, according to (18), equation $k_z d = m\pi$ together with $k_{0z}b = n\pi$ determines the band contact points. So it is easy to conclude, that the periodic structure becomes an absolutely transparent for the plane wave, independent of its polarization at the band contact points. The obtained result does not contradict the well-known Bragg reflection condition at the band boarders at all. It is necessary to note at once, that the contact point cannot be considered as a boarder point for any allowed band. At the contact we have one large allowed band, formed out from the two allowed bands, for which the contact point is interior.

An analogous complete transparence is well known, but for $p$ waves only, when the incident angle is that of Brewster [15]. In this case the reflection amplitude for $p$ waves vanishes ($r^p = 0$). The fact, that under Brewster`s condition the allowed bands contact easily, may be shown from the equations (10), (11), which define the band boarders for $p$ waves. Indeed, equations (10) and (11) have solutions, which coincide with each other not only at the condition (18), but also at the following condition

$$\frac{\varepsilon k_{0z}}{\varepsilon_0 k_z} = \frac{\varepsilon_0 k_z}{\varepsilon k_{0z}}. \tag{22}$$

Using Snell's law it is easy to see, that (22) is nothing else, than Brewster law ($tg\alpha = \sqrt{\varepsilon/\varepsilon_0}$).

Note, that in the case of Brewster angle, the periodic structure is transparent for $p$ wave at an arbitrary frequency. For the case considered here, the structure is transparent for an arbitrary polarization, but at a definite value of incidence angle the wave must have only a definite frequency.

## Conclusion

In the present paper the problem of reflection and transmission of a plane electromagnetic wave for an one-dimensional periodic layered structure is considered. The transmission band locations dependence on the problem parameters are considered both for $s$ and $p$ polarized wave. Using the standard theory, the equations, defining the even and odd band boarders, are obtained separately. The graphs for the band location dependence on various structural and material parameters of layered structure are brought. It is proved that at a certain choice of the problem parameters the contact of allowed bands is possible. An analytical condition is obtained, which defines the possibility of such contact. This condition has the same form both for two polarizations. Numerical calculation are brought, which show that the band width dependence on the structure parameters is not a monotonous function. This non-monotony appears at the band contact regions.

The transmission coefficient behavior is considered at the contact points. It is shown that at these points the periodic structure becomes transparent for an arbitrary polarized wave, at its corresponding frequency.

## Acknowledgment

I would like to express my gratitude to D. Sedrakian, H. Eritsyan, A. Gevorgyan and L. Mouradian for the discussion of the problem results.

Figure 1. The location of the transmission bands (darkened regions) depending on dielectric constant of the layer at normal incidence of wave in case of eight different values of $b/d = 0.5, 1, 1.5, 2, 2.5, 3, 3.5, 4$ parameter (see figures from a) to h), correspondingly).

Figure 2. The location of the transmission bands (darkened regions) at normal incidence of wave depending on dimensionless parameter $d/b$ in case of eight different values of $\varepsilon = 1.5, 3, 4.5, 6, 7.5, 9, 10.5, 12$ (see figures from a) to h), correspondingly).

Figure 3. The location of the transmission bands (darkened regions) for $s$ (figures a), c), e), g)) and $p$ (figures b), d), f), h)) waves at $\varepsilon = 7$ in case of four different values of angle of incidence $\alpha = 20^0, 40^0, 60^0, 80^0$ depending on dimensionless parameter $d/b$.

Figure 4. The location of the transmission bands (darkened regions) for $s$ wave at two different values of $b/d$ in case of four different angles of incidence. Graphs, given in pictures a), b), c), d) correspond to value $b/d = 0.5$ at angles of incidence $\alpha = 40^0, 50^0, 60^0, 70^0$. Graphs, given in pictures e), f), g), h) correspond to the value $b/d = 1$ at $\alpha = 50^0, 60^0, 70^0, 80^0$.

Figure 5. The location of the transmission bands (darkened regions) for $p$ wave at two different values of $b/d$ in case of four different angles of incidence. Graphs, given in pictures a), b), c), d) correspond to value a), b), c), d) correspond to the value $b/d = 0.5$ at angles of incidence $\alpha = 30^0, 40^0, 50^0, 60^0$. Graphs, given in pictures e), f), g), h) correspond to the value $b/d = 1$ at $\alpha = 40^0, 50^0, 60^0, 70^0$.

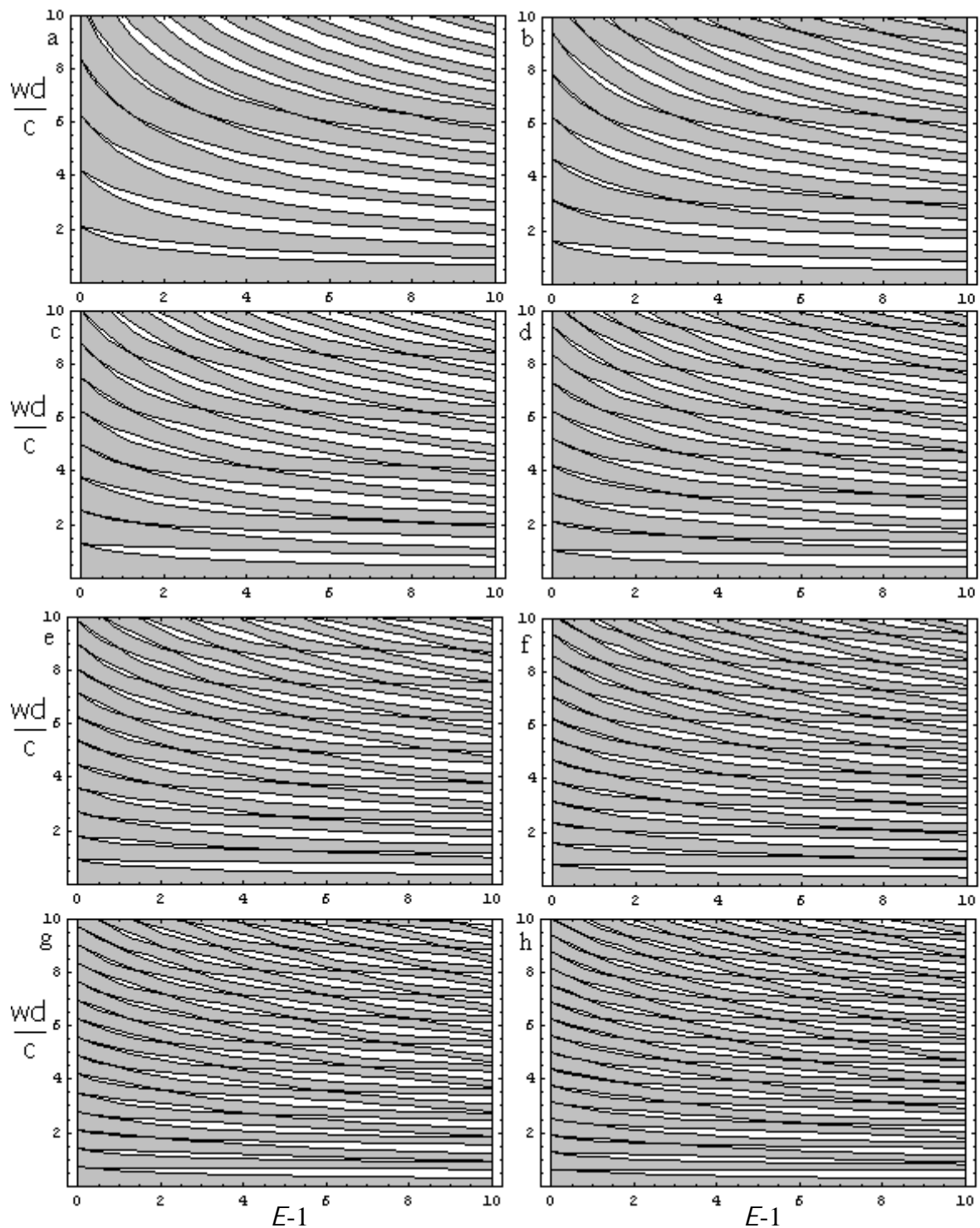

Figure 1. The location of the transmission bands (darkened regions) depending on dielectric constant of the layer at normal incidence of wave in case of eight different values of $b/d = 0.5, 1, 1.5, 2, 2.5, 3, 3.5, 4$ parameter (see figures from a) to h), correspondingly).

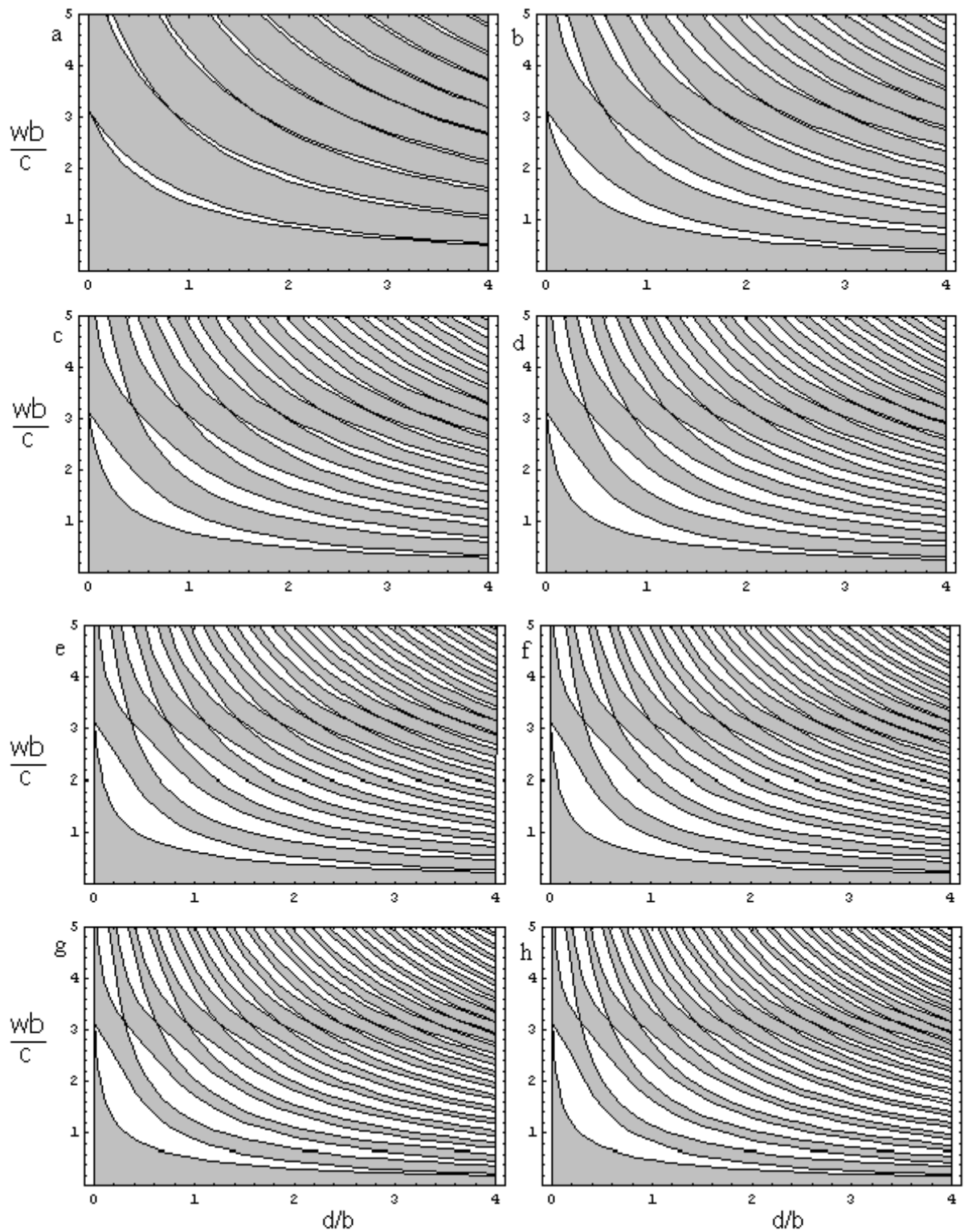

Figure 2. The location of the transmission bands (darkened regions) at normal incidence of wave depending on dimensionless parameter $d/b$ in case of eight different values of $\varepsilon = 1.5, 3, 4.5, 6, 7.5, 9, 10.5, 12$ (see figures from a) to h), correspondingly).

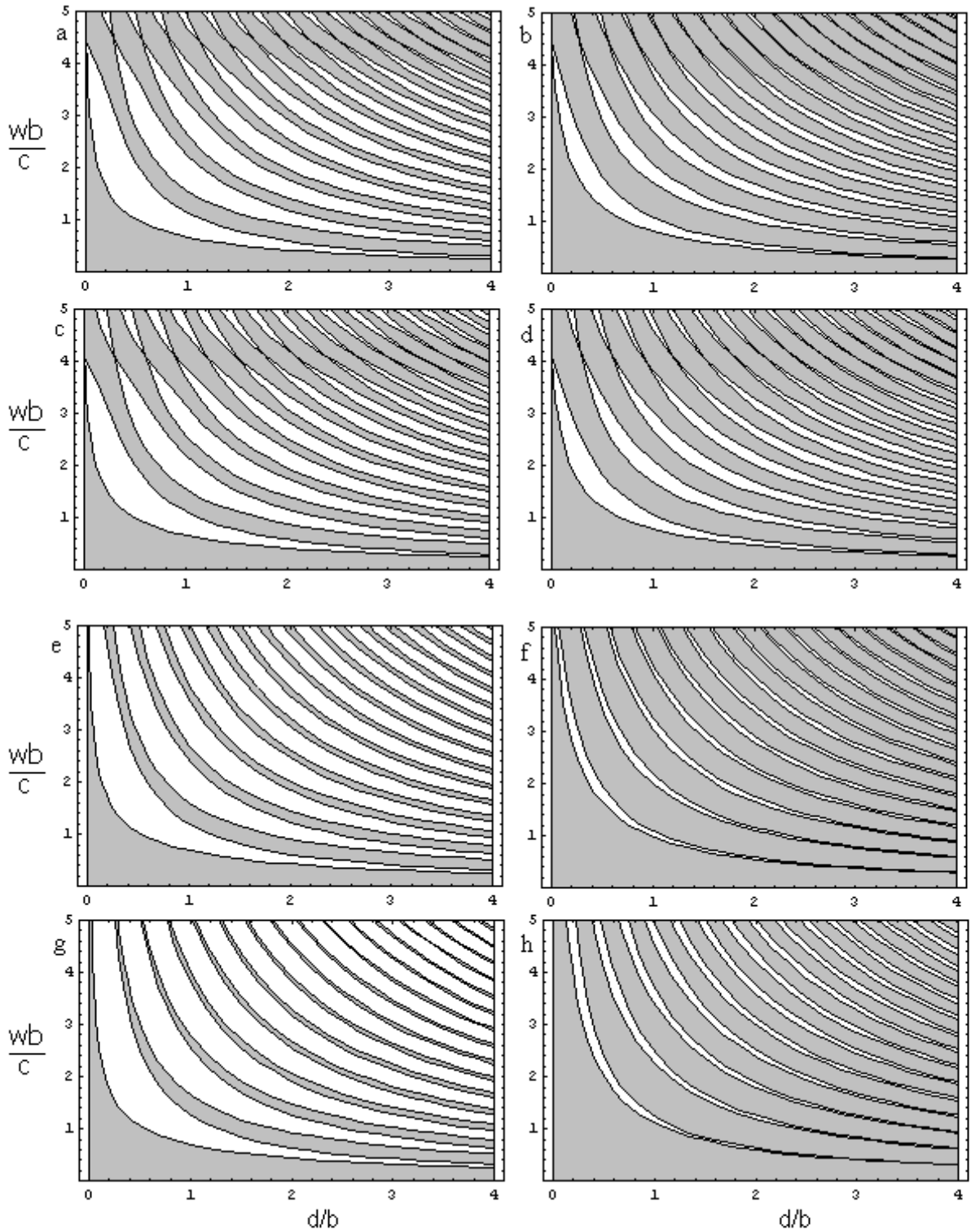

Figure 3. The location of the transmission bands (darkened regions) for $s$ (figures a), c), e), g)) and $p$ (figures b), d), f), h)) waves at $\varepsilon = 7$ in case of four different values of angle of incidence $\alpha = 20^0, 40^0, 60^0, 80^0$ depending on dimensionless parameter $d/b$.

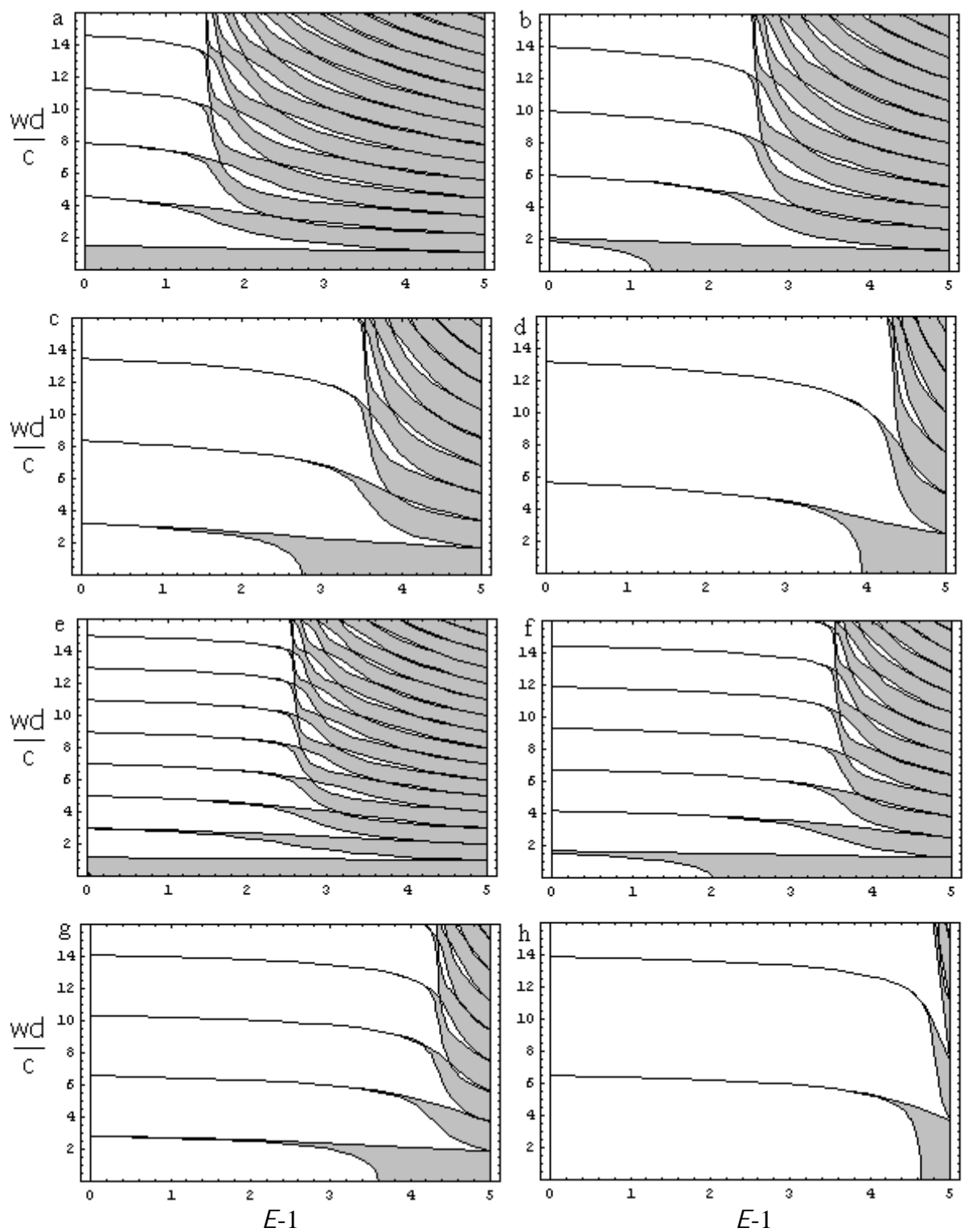

Figure 4. The location of the transmission bands (darkened regions) for $s$ wave at two different values of $b/d$ in case of four different angles of incidence. Graphs, given in pictures a), b), c), d) correspond to value $b/d = 0.5$ at angles of incidence $\alpha = 40^0, 50^0, 60^0, 70^0$. Graphs, given in pictures e), f), g), h) correspond to the value $b/d = 1$ at $\alpha = 50^0, 60^0, 70^0, 80^0$.

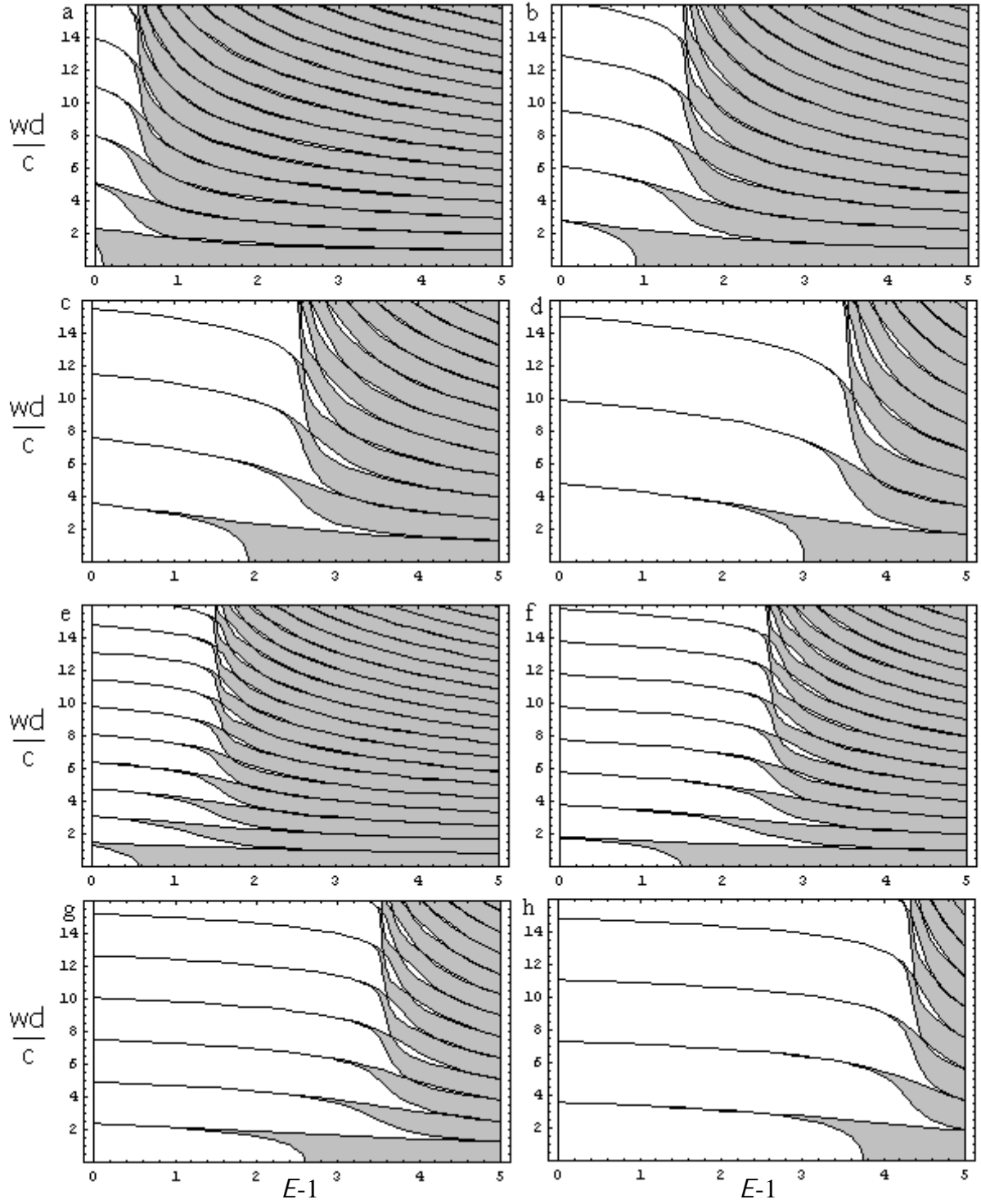

Figure 5. The location of the transmission bands (darkened regions) for *p* wave at two different values of $b/d$ in case of four different angles of incidence. Graphs, given in pictures a), b), c), d) correspond to value a), b), c), d) correspond to the value $b/d = 0.5$ at angles of incidence $\alpha = 30^0$, $40^0, 50^0, 60^0$. Graphs, given in pictures e), f), g), h) correspond to the value $b/d = 1$ at $\alpha = 40^0$, $50^0, 60^0, 70^0$.